\documentclass[a4paper]{article}
\pdfoutput=1

\usepackage[utf8]{inputenc}
\usepackage[T1]{fontenc}
\usepackage[english]{babel}

\usepackage[pdftex,breaklinks=true,bookmarks=true,pdfborder={0 0 0},colorlinks=false]{hyperref}

\usepackage{listings}
\usepackage{amsfonts}
\usepackage{amsmath}
\usepackage{cleveref}
\usepackage{graphicx}
\usepackage[caption=false]{subfig}

\newcommand{\ie}{i.e.~}
\newcommand{\eg}{e.g.~}
\newcommand{\cf}{cf.~}
\newcommand{\vs}{vs.~}

\usepackage{enumitem}
\setlist{nolistsep}
\usepackage{multirow}
\usepackage{array}

\begin{document}

\author{Raphael ‘kena’ Poss\\University of Amsterdam, The Netherlands}
\title{Characterizing traits of coordination}

\maketitle

\begin{abstract}
  How can one recognize coordination languages and technologies? As
  this report shows, the common approach that contrasts coordination
  with computation is intellectually unsound: depending on the
  selected understanding of the word ``computation'', it either
  captures too many or too few programming languages. Instead, we
  argue for objective criteria that can be used to evaluate how well
  programming technologies offer coordination services. Of the various
  criteria commonly used in this community, we are able to isolate
  three that are strongly characterizing: black-box componentization,
  which we had identified previously, but also interface extensibility
  and customizability of run-time optimization goals. These criteria
  are well matched by Intel's Concurrent Collections and AstraKahn,
  and also by OpenCL, POSIX and VMWare ESX.
\end{abstract}


\setcounter{tocdepth}{1}
\tableofcontents

\clearpage

\section{Introduction}\label{sec:intro}

The author of this report studies a research community whose
specialization is the management of software components in
multi-component applications. The members of this community have
agreed on a common linguistic referent for their activities in this
field: the word ``\emph{coordination}''.

The main output from this community is a combination of programming
languages and operating software aimed at optimizing the run-time
execution of applications built by hierarchical composition of
components. Example technologies whose authors self-identify as
``working on coordination'' include
S-NET~\cite{shafarenko.09.apc,grelck.08.ppl}, AstraKahn~\cite{shafarenko.13.ak} and Intel's Concurrent
Collections (CnC)~\cite{knobe.09,budimlic.11}.

A recurring theme in the discussions within this community and with
external observers is whether and how much coordination differs from
other forms of programming. This topic is usually introduced with
either of two questions: ``what is coordination exactly?'' and ``what
distinguishes research on coordination from other research on
programming language design and implementation?''

As it happens, different answers are used in these conversations
depending on who is asking, who is answering and the topic at
hand. This author has observed a consensus in the community that these
answers are all \emph{accepted} by the researchers as valid
descriptions of their line of work.

Of these explanations, we can recognize four groups:
\begin{itemize}
\item \emph{self-referential} explanations: a research activity is
  considered related to ``coordination'' if it self-identifies as
  such. For example, ``this language is a coordination language because its
  designers call it a coordination language'';
\item \emph{negative space} explanations: an existing field of study is
  selected \textsl{ad hoc}, then a research
  activity is considered related to ``coordination'' if it
  self-identifies as ``not related to'' the selected research. For
  example, ``this language is a coordination language because its
  designers do not focus on software modeling'' (or functional
  programming, or model checking, etc.);
\item \emph{void} explanations: a word is selected \textsl{ad hoc}
  with no well-defined meaning, then a research activity is considered
  related to ``coordination'' if it self-identifies as ``not related
  to'' the selected word. For example, ``this language is a
  coordination language because its designers do not intend it to be a
  computation language'' without a clear definition for the word ``computation'' (\cf \cref{sec:comp});
\item explanations \emph{by qualification}: some well-defined, objective,
  observer-in\-de\-pen\-dent criteria on programming languages and operating
  software are identified, then ``coordination'' is defined based on the criteria. For
  example, ``this language is a coordination language because it
  offers facilities to assemble applications from black-box
  components'', together with a careful definition of ``black-box component'', constitutes a qualified explanation.
\end{itemize}

The self-referential, negative space and void explanations are, by
construction, factually vacuous: a newcomer audience exposed to them
will not learn anything about \emph{what the researcher using the
  explanation actually does in their work}. At best, the audience
may understand that the researcher needs a keyword to motivate
specialized attention and funding, but not more. These forms of
explanations are thus not further considered here.

Instead, this report reviews the criteria where consensus exists in
the community that they can be used to recognize coordination objectively.

The discussion is presented in two parts. In \cref{sec:crit}, we
identify and detail the objective criteria that have been previously
named and used by members of the research community. We then examine in \cref{sec:eval}
how well this community's technology matches their self-selected
criteria, and also how well other technologies match the same
criteria. We then use this analysis to isolate which criteria most
strongly characterize the work of these researchers. Separately, in
\cref{sec:comp} we examine the arguments that oppose ``coordination''
to ``computation'', and we analyze how much
objective understanding can be extracted from them. We then
conclude in \cref{sec:conc}.

\section{Qualifying criteria}\label{sec:crit}

We reuse below the definition of ``component'' and component-based
design from \cite{batory.92,poss.13.coord}: components are defined by
their interface, which specifies how they can be used in applications,
and one or more implementations which define their actual behavior.
The two general principles of component-based design are then phrased
as follows. The first is interface-based integration: when a designer
uses a component for an application, he agrees to only assume what
is guaranteed from the interface, so that another implementation can
be substituted if needed without changing the rest of the
application. The second is reusability: once a component is
implemented, a designer can reuse the component in multiple
applications without changing the component itself.

Based on this definition, the word ``coordination'' is only used in
the context of languages and infrastructures that enable
component-based design\footnotemark{}.
\footnotetext{arguably, most contemporary
programming technologies already enable component-based design; however we explicitly state the requirement to clearly scope the discussion.}

\paragraph{Separable provisioning (Sp):} the language and its infrastructure
enable the reuse of components provided by physically separate
programmers, and where the considered technology is the only
communication means between these providers. For example, a technology
that offers the ability to build a specification from different files
matches this criterion.

\paragraph{Interface extensibility (Ie):} the infrastructure enables an
application designer to redefine and extend component interfaces
independently from component providers, and extended interfaces can
influence execution. For example, a technology that offers the ability
to annotate a component to indicate  \textsl{post hoc} that it is functionally pure (without
state and deterministic), \eg via a pragma or
metadata, and which can exploit this annotation to increase execution
parallelism, matches this criterion.

\paragraph{Separable scheduling (Ss):} the programmer can delegate to
the technology the responsibility of choosing when (time) and where
(space/resources) to execute concurrent component activities.  A
different but equivalent phrasing for the same criterion is the
ability given to a programmer to define a partial scheduling order
between component activities, and ability given to the technology to
decide arbitrary actual schedules as long as the partial order is
respected.  For example, a language where programmers can declare a
data-parallel operation and where the infrastructure decides how to
schedule the operation matches this criterion.

\paragraph{Adaptable optimization (Ao):} the technology provides run-time
optimization mechanisms that can adapt to different execution
environments without changing the application specification. For
example, a technology which can decide different placements when faced
with different amounts of parallelism in hardware matches this
criterion, and so does a technology able to decide different schedules
over time when faced with different constraints on data locality (\eg
cache sizes).

\paragraph{Customizable optimization goals (Cog):} the application
designer can specify different optimization goals at run-time (or no
earlier than when the specification work has completed) and the
technology chooses different execution strategies based on them. For
example, a technology which enables to select between ``run fast'' and
``use less memory'' during execution matches this criterion.

\paragraph{Black-box componentization (Bb):} the application
designer can specify an application using components only known
to the technology by name and interface, and the technology provides a
run-time interfacing mechanism previously agreed upon with component
providers to integrate the components. For example, a technology
which can link component codes compiled from different programming languages
without requiring link-time cross-optimization matches this criterion.
This is the main criterion proposed in~\cite{poss.13.coord}.

\paragraph{Exploitable Turing-incompleteness (Eti):} the specification
language is not Turing complete but can still be used to define
interesting / useful applications. For example, a technology whose
advertised specification language can only define static acyclic data
flow graphs of components matches this criterion.

\section{Criteria evaluation}\label{sec:eval}

\begin{table}[t]
\scriptsize
\hspace{-3em}\begin{tabular}{|l|>{\raggedright\arraybackslash}p{.16\textwidth}l|ccccccc|l|l|}
\hline
\multicolumn{3}{|l|}{Technology} & \multicolumn{7}{l|}{Criteria} &
\multicolumn{2}{l|}{Scores} \\
\cline{2-12}
 & Variant              & Granularity & Sp & Ie & Ss & Ao & Cog & Bb & Eti & Total & I only \\

\hline
\hline
\multirow{3}{*}{S-NET \cite{shafarenko.09.apc,grelck.08.ppl}}
 &                      & boxes       & I  &    & I  & I  &     & I  &     & 4     & 4 \\
 &                      & networks    &    &    & I  & I  &     &    &     & 2     & 2 \\
 & no filters / star    & boxes       & I  &    & I  & I  &     & I  & E   & 5     & 4 \\

\hline
\multirow{3}{*}{S+NET \cite{poss.13.apc,poss.13.spnet}}
 &                      & boxes       & I  &    & I  & I  & I   & I  &     & 5     & 5 \\
 &                      & networks    & I  &    & I  & I  & I   &    &     & 4     & 4 \\
 & no trans. / star     & boxes       & I  &    & I  & I  & I   & I  & E   & 6 $\leftarrow$    & 5 \\

\hline
\multirow{2}{*}{AstraKahn \cite{shafarenko.13.ak}}
 &                      & boxes       & I  & I  & I  & I  & I   & I  &     & 6 $\leftarrow$    & 6 $\leftarrow$ \\
 &                      & networks    &    &    & I  & I  & I   &    &     & 3     & 3 \\

\hline
CnC \cite{knobe.09,budimlic.11} &                   & steps       & I  & I  & I  & I  & I   & I  &     & 6 $\leftarrow$    & 6 $\leftarrow$ \\

\hline
\hline
\multirow{7}{*}{C \cite{isoc99,isoc11,openmp,opencl,ieeeposix.08}}
 & freestanding         & functions   & I  &    &    &    &     & I  & E   & 3     & 2 \\
 & freestanding         & statements  & I  &    & E  & E  &     &    & E   & 4     & 1 \\
 & freestanding         & expressions & I  & I  & E  & E  &     &    & E   & 5     & 2 \\
 & OpenMP               & statements  & I  & I  & I  & I  &     &    & E   & 5     & 4 \\
 & OpenCL               & kernels     & I  & I  & I  & E  & E   & I  &     & 6 $\leftarrow$    & 4 \\
 & ISO11, hosted        & threads     & I  &    & I  & E  &     & I  &     & 4     & 3 \\
 & ISO99, POSIX         & threads     & I  & E  & I  & E  & E   & I  &     & 6 $\leftarrow$    & 3 \\
 & ISO99, POSIX         & processes   & I  & I  & I  & I  & I   & I  &     & 6 $\leftarrow$    & 6 $\leftarrow$ \\

\hline
C++ \cite{isocpp11} & ISO11, POSIX      & classes     & I  & I  & E  & E  & E   &    &     & 5     & 2 \\

\hline
\multirow{2}{*}{Haskell \cite{marlow.09}}
 &  GHC                 & functions   & I  & I  & I  & I  &     & I  &     & 5     & 5 \\
 &  GHC                 & packages    & I  & I  & I  & I  &     &    &     & 4     & 4 \\

\hline
\multirow{2}{*}{SAC \cite{grelck.06.ijpp}}
 &                      & functions   & I  & I  & I  & I  &     & I  &     & 5     & 5 \\
 &                      & modules     & I  &    & I  & I  &     &    &     & 3     & 3 \\

\hline
Unix shell \cite{ieeeposix.08}
& POSIX                 & processes   & I  & I  & I  & I  & I   & I  &     & 6 $\leftarrow$    & 6 $\leftarrow$ \\

\hline
VMWare ESX
&                & virtual machines   & I  & I  & I  & I  & I   & I  &     & 6 $\leftarrow$    & 6 $\leftarrow$ \\

\hline
\end{tabular}
\caption{How various technologies match the proposed criteria.}\label{tab:eval}
\end{table}

We evaluate in \cref{tab:eval} how much different technologies match
the criteria defined above: the criterion are listed in columns, the
technologies in rows, each intersection states whether the technology
matches the criterion, and a score column at the right side sums the
number of criterion matched. Arrows in the score columns indicate the
rows with highest scores.

We review both technologies that self-identify as ``coordination'',
including S-NET and CnC named previously, and other technologies that do
not identify as such: various C and C++ implementations, Glasgow
Haskell, Single-Assignment C
(SAC), the standard Unix shell in a POSIX environment and VMWare ESX.

While constructing \cref{tab:eval}, we highlighted the
following:
\begin{itemize}
\item \emph{granularity}: each technology may offer multiple levels of
  component granularities, and may not match the same criteria
  depending on the granularity considered. For example, the C language
  offers black-box componentization for entire functions but not for
  individual statements. To reflect this, multiple rows with different
  granularities are used for each technology in the table.
\item \emph{intent}: a technology may happen to match a criterion
  although this match was not primarily intended by its designers. For
  example, a freestanding implementation of the C language (without
  library) happens to be Turing incomplete and still quite useful,
  although this was arguably not intended by its designer (nor
  commonly known of its users). To reflect this, we use the letters
  ``I'' (by intent) and ``E'' (emergent) at each intersection and
  provide two score columns in the right side.
\end{itemize}

From this first evaluation table we observe the following.

First, separable provisioning (Sp) is generally prevalent. Although it is a
prerequisite to component-based design and thus coordination, its
availability in a particular technology does not predict its score in
our table. Therefore, it is a poor criterion to characterize coordination.

Similarly, separable scheduling (Ss) and adaptable optimization (Ao)
are also relatively prevalent. Although the benefits of separate
scheduling and adaptable optimization wrt. performance speedups on parallel hardware is often
used to highlight the benefits of coordination, other technologies
which do not self-identify as ``coordination'' (\eg Haskell, OpenCL,
SAC) also exhibit these features and can reap their associated
benefits. These criteria may thus be phrased as
``prerequisites'' to recognize coordination but they are not
characterizing.

Also, the ``exploitable Turing-incompleteness'' (Eti) criterion is,
perhaps surprisingly, difficult to match. The main reason, which we
outline in \cref{sec:comp}, is that it is actually quite difficult to
design a programming language which is not Turing-equivalent.

Finally, the table reveals that none of the proposed criteria clearly
separates technology that self-identify as ``coordinating'' from those
that don't. The evaluation of whether a technology can be considered
as coordination cannot yield a boolean value and instead lies on a
spectrum of ``more or less able to coordinate''.

\begin{table}[t]
\centering
\scriptsize
\begin{tabular}{|l|>{\raggedright\arraybackslash}p{.16\textwidth}l|ccc|l|l|}
\hline
\multicolumn{3}{|l|}{Technology} & \multicolumn{3}{l|}{Criteria} &
\multicolumn{2}{l|}{Scores} \\
\cline{2-8}
 & Variant              & Granularity & Ie & Cog & Bb & Total & I only \\

\hline
\hline
\multirow{2}{*}{S-NET}
 &                      & boxes       &    &     & I  & 1     & 1 \\
 &                      & networks    &    &     &    & 0     & 0 \\

\hline
\multirow{2}{*}{S+NET}
 &                      & boxes       &    & I   & I  & 2     & 2 \\
 &                      & networks    &    & I   &    & 1     & 1 \\

\hline
\multirow{2}{*}{AstraKahn}
 &                      & boxes       & I  & I   & I  & 3  $\leftarrow$    & 3  $\leftarrow$ \\
 &                      & networks    &    & I   &    & 1     & 1 \\

\hline
CnC &                   & steps       & I  & I   & I  & 3  $\leftarrow$    & 3  $\leftarrow$\\

\hline
\hline
\multirow{7}{*}{C}
 & freestanding         & functions   &    &     & I  & 1     & 1 \\
 & freestanding         & statements  &    &     &    & 0     & 0 \\
 & freestanding         & expressions & I  &     &    & 1     & 1 \\
 & OpenMP               & statements  & I  &     &    & 1     & 1 \\
 & OpenCL               & kernels     & I  & E   & I  & 3  $\leftarrow$    & 2 \\
 & ISO11, hosted        & threads     &    &     & I  & 1     & 1 \\
 & ISO99, POSIX         & threads     & E  & E   & I  & 3  $\leftarrow$    & 1 \\
 & ISO99, POSIX         & processes   & I  & I   & I  & 3  $\leftarrow$    & 3  $\leftarrow$\\

\hline
C++ & ISO11, POSIX      & classes     & I  & E   &    & 2     & 1 \\

\hline
\multirow{2}{*}{Haskell}
 &  GHC                 & functions   & I  &     & I  & 2     & 2 \\
 &  GHC                 & packages    & I  &     &    & 1     & 1 \\

\hline
\multirow{2}{*}{SAC}
 &                      & functions   & I  &     & I  & 2     & 2 \\
 &                      & modules     &    &     &    & 0     & 0 \\

\hline
Unix shell & POSIX      & processes   & I  & I   & I  & 3  $\leftarrow$    & 3  $\leftarrow$\\

\hline
VMWare ESX &      & virtual machines  & I  & I   & I  & 3  $\leftarrow$    & 3  $\leftarrow$\\

\hline
\end{tabular}
\caption{How various technologies match the proposed criteria (simplified).}\label{tab:eval2}
\end{table}

From these observations, we can select the criteria most strongly
matched by these technologies that the researchers would like to
objectively describe as ``strongly coordinating.'' This suggests the criteria Ie, Cog and Bb and the summary in \cref{tab:eval2}. As
the table shows, AstraKahn, CnC, OpenCL, POSIX and VMWare ESX can be
considered strongly coordinating, each at their preferred component
granularity: boxes, steps, kernels, threads/processes and virtual
machines, respectively.

\section{Problems of ``coordination \vs computation''}\label{sec:comp}

During the discussions around coordination, this author has observed a
prevalent use of the following arguments by the members of the community:
\begin{enumerate}\bfseries
\item ``coordination technologies can be distinguished from computation technologies'';
\item ``what differentiates coordination and computation technologies
  is the intent of the designer: the designers of coordination
  languages do not focus on computation'';
\item ``there exist `pure' coordination languages that cannot be used to
  specify computations.''
\end{enumerate}

All three arguments are motivated by a subjective, human desire of the
involved individuals, that to create a ``us-versus-them'' vision of
the research. The ulterior motive is to generate specialized attention
and attract dedicated funding. In fairness to this community, we
highlight here that this ulterior motive is shared by most academic
researchers regardless of their field of study.

However, despite and regardless of the motive and its subjectivity,
the individuals involved claim (both implicitly and explicitly) that
these three arguments can be recognized as objective by an external
observer, \ie they can stand and be defended at face value.

What interests us here is that all three arguments require some shared
understanding of what is meant by ``computation.'' If no shared
understanding can be found, then all three arguments are void and thus
intellectually irrelevant.

Moreover, if a shared understanding can be found, then only argument
\#3 is objectively qualified. Even with a shared understanding of
computation, arguments \#1 and \#2 remain at best ``negative space''
arguments (\cf \cref{sec:intro}) and still do not inform about what
coordination actually entails.

To see how much of argument \#3 can be ``saved'' for the purpose of
objective discussions, we need to investigate two points. The first is
how much shared understanding can be gathered around the term
``computation''. The second is whether, assuming some shared
understanding of what ``computation'' entails, argument \#3 actually holds:
that languages that cannot express computation actually exist, and can
be called coordination languages.

\subsection{About the notion of computation}

As of this writing, there exists no formal definition of what
constitutes a computation in general. What is known empirically is
that for any function of mathematics it is often possible to build a
machine which can calculate the value of this function for some
input. What is known formally, is that for any given number function
of mathematics it is always possible (in theory) to build a machine
that can calculate the value of this function. What is not known
however, is the set of all mathematical functions a given
concrete (real-world) machine can reproduce; and whether it is
possible to build a machine for all possible mathematical functions,
not only number functions. Meanwhile, people can be observed to also
build machines to perform work that is not described formally but is
still considered useful.

In this context, two approaches can be taken to define
``computation''. One can seek formalism at all costs, and restrict the
shared understanding to Church and Turing's \emph{thesis}: that the
set of computations is exactly the set of all possible input-output
transformations by any theoretical Turing machine.  However, this
Manichean approach excludes a range of machine activities that are
commonly considered to be ``computations'' in practice, too:
transformation and communication of real (physical) variables,
non-deterministic operations over parallel hardware with loosely
synchronized time, ongoing processes without a start event, etc.

The other way to define ``computation'' is to identify some useful
real-world artefacts and behaviors, call them ``computation'' axiomatically, then
reverse-engineer which languages and formal systems can be used to
specify them. There are multiple ways to do so; here are the two
such definitions that seem to gather most consensus:

\begin{itemize}
\item \emph{``terminating value computations''}: any operation which
  consumes \emph{a finite supply of static data} as input, runs for a
  \emph{finite amount of time} and produces \emph{a finite supply of
    static data} as output. This includes but is not limited to the
  observable behavior of halting Turing machines;
\item \emph{``process computations''}: any operation which is running
  within \emph{a well-formed space boundary} (\eg a specific component
  of a machine), running at a \emph{measurable cost} and that is
  \emph{controllable}: where an external agent (\eg a person or
  another system) can start, stop, observe, accelerate, slow (etc.)
  the operation.
\end{itemize}

The choice of approach also defines the objective substance of any
discussion that capitalizes on the notion of computation. Different
choices result in different, possibly conflicting
understandings. Therefore, any situation where the word
``computation'' is used casually to support negative space arguments
should be reviewed with critical care; in particular, one should feel challenged
to isolate and clarify explicitly what assumptions are being made.

\subsection{Languages that ``cannot specify computations''}

There are two interpretations for the phrase ``cannot specify
computations'': either ``cannot specify \emph{any} computation'' or
``cannot specify \emph{all} computations''. The argument ``There
exists pure coordination languages that cannot specify computation''
thus defines two classes of languages: \emph{computation-less}
languages which cannot be used to define any computation whatsoever;
and \emph{incomplete} languages which can only be used to specify a
limited subset of computations.

\begin{table}[t]
\footnotesize
\hspace{-.1\textwidth}\begin{tabular}{|p{.4\textwidth}|p{.4\textwidth}p{.4\textwidth}|}
\hline
Chosen definition for ``computation'' & Computation-less languages & Incomplete computation languages \\
\hline
Behavior of Turing machines & (none known) & few languages, but includes C \\
Terminating value computations & (none known) & most languages \\
Process computations & some declarative languages, including Prolog, pure $\lambda$-calculus, HTML & most languages \\
\hline
\end{tabular}
\caption{Languages that cannot specify computations.}\label{tab:compt}
\end{table}

Both can only be discussed in the context of a specific, \textsl{a
  priori} chosen understanding of the word ``computation'' as
described in the previous section. We collate in \cref{tab:compt} a
condensed inventory of existing programming languages that are either
computation-less or incomplete for the various definitions of
``computation'' isolated previously.

\Cref{tab:compt} enables three observations.

The first is that it is difficult to find concrete computation-less
languages, for any definition of ``computation''. In general, it is
actually difficult to design a computation-less language: any language
that is able to define a dynamic evaluation that can react to state,
regardless of how dynamic its input is, can be tricked at a
higher-level to define \emph{some} computations. For example, with
S-NET one can define operations using Peano arithmetic on the depth of
the run-time expansion of a ``star'' combinator over a synchrocell,
using only record types to perform choices. A computation-less
language should either prevent its user from defining a dynamic
evaluation, or restrict the evaluation to be state-insensitive (or
both). It is debatable whether languages with such restrictions can be
called ``programming'' languages at all.

The second is that if we consider process computations in general and
we understand that ``pure coordination languages are those languages that are
computation-less'', we would need to accept languages like Prolog,
$\lambda$-calculus or HTML as coordination languages. This does not
appear compatible with the vision of the coordination community being
studied.

The third is that there are ``too many'' languages that are incomplete
with regards to each definition of ``computation'' to hold a strong
us-versus-them argument. For the two informal definitions, \ie
terminating value computations and process computations, if we
understand that ``pure coordination languages are those languages that
are incomplete with regard to specifying computation'', then virtually
any programming language in use today is a coordination language.  If
we take the formal definition instead (Turing-incompleteness), then C
would also qualify as a coordination language because C is also
Turing-incomplete\footnote{we consider here the C language without its
  standard library as defined in \cite{isoc99,isoc11}. This author can
  provide a demonstration of C's Turing-incompleteness upon
  request.}. Again, this does not appear compatible with the vision of
this community.

To summarize, it may not be possible to use argument \#3 successfully
to motivate specialized attention to the work of this community.

\section{Conclusion}\label{sec:conc}

We have reviewed in this report the commonly used, subjective argument
that ``coordination can be contrasted to computation''. We have
revealed that this argument and all currently used related phrasings
are largely intellectually unsound and we conclude they cannot be used
to support specialized scientific attention towards ``coordination''
as a research activity.

Instead, we have highlighted that research on ``coordination'' can be
supported objectively using motivating arguments based on objective
criteria. Of the various candidate criteria that have been proposed so
far, we have shown that only three characterize the work of the
researchers involved:

\begin{itemize}
\item \emph{interface extensibility}: the ability to extend or replace component interfaces
  arbitrarily after components are provided, and define valid
  composite behavior using the modified interfaces even if they
  conflict with the internal structure of the components;
\item \emph{customizable optimization goals}: the ability to specify
  different optimization goals after the application has been
  specified, \eg during execution, and the ability of the technology
  to use different execution strategies to match the custom goals;
\item \emph{black-box componentization}: the ability to specify composite
  applications from components only known by name and interface, and
  the existence of run-time interfacing mechanisms that do not require the
  coordination technology to know anything about the internal structure of components.
\end{itemize}

Of these three criteria, we had previously~\cite{poss.13.coord}
identified the last as a clear objective criterion to recognize
coordination, and we had recognized that programming technologies are
``more or less coordinating'' depending on how well they match the
criterion. In the present report, we have extended this argument to
the other two criteria, and recognized several concrete
coordination technologies: AstraKahn and Intel's CnC, but also OpenCL,
POSIX and VMWare ESX.

\section*{Acknowledgements}
\addcontentsline{toc}{section}{Acknowledgements}

This document reports on thoughts nurtured during successive
discussions with Merijn Verstraaten, Alex Shafarenko, Sven-Bodo
Scholz, Kath Knobe and Clemens Grelck.

\newcommand{\etalchar}[1]{#1} 
\addcontentsline{toc}{section}{References}
\bibliographystyle{is-plainurl}
\bibliography{doc}

\end{document}